\documentclass[prl,showpacs]{revtex4}

\usepackage{graphicx}
\usepackage{amsmath,amstext,bm}
\usepackage{amsfonts}

\renewcommand{\vec}{\bm}

\begin{document}


\title{Ultrafast demagnetization of ferromagnetic transition metals:\\
The role of the Coulomb interaction}

\author{Michael Krau\ss}
\author{Tobias Roth}
\author{Sabine Alebrand}
\author{Daniel Steil}
\author{Mirko Cinchetti}
\author{Martin Aeschlimann}
\author{Hans Christian Schneider}
\affiliation{Department of Physics and
  Research Center OPTIMAS, University of Kaiserslautern,
  Erwin-Schr\"odinger-Str. 46, 67653 Kaiserslautern, Germany}  

\date{\today}

\pacs{75.10.Lp,71.70.Ej,78.47.-p,78.47.J-}

\begin{abstract}
The Elliott-Yafet (EY) mechanism is arguably the most promising
candidate to explain the ultrafast demagnetization dynamics in
ferromagnetic transition metals on timescales on the order of
100\,fs. So far, only electron-phonon scattering has been analyzed as
the scattering process needed to account for the demagnetization due
to the EY mechanism. We show for the first time that the
\emph{electron-electron} scattering contribution to the EY mechanism
has the potential to explain time-resolved magneto-optical Kerr effect
measurements on thin magnetic Co and Ni films, without reference to a
phononic ``spin bath.''
\end{abstract}

\maketitle

Current research in femtosecond magnetism is concerned with
elucidating the fundamental mechanisms of light-induced spin dynamics
as well as searching for potential applications in data
processing~\cite{vaterlaus,beaurepaire,rhie}. Despite of important
experimental studies employing various time-dependent magneto-optical
techniques, no consensus on a microscopic understanding of ultrafast
magnetization dynamics in ferromagnets has emerged. Rather,
demagnetization dynamics is typically described in the framework of
the phenomenological three-temperature model. In this model,
temperatures are assigned to the electron, lattice and spin
``subsystems,'' and the exchange of energy (and spin) is driven by the
temperature differences between the respective subsystems. Although
the three-temperature model provides an intuitive picture of
demagnetization, its relation to the microscopic dynamics behind the
demagnetization is still an active field of research.

The most popular candidate~\cite{koopmans:prl05} for the microscopic
process behind ultrafast demagnetization is a mechanism of the
Elliott-Yafet (EY) type~\cite{zutic:review}. In the EY mechanism, the
demagnetization arises because, in the presence of the spin-orbit
interaction, spin is not a good quantum number, so that \emph{any}
momentum-dependent scattering mechanism changes the spin admixture
when an electron is scattered from state $|\vec{k}\rangle$ to
$|\vec{k}+\vec{q}\rangle$. So far, the scattering process responsible
for the EY mechanism has been assumed to be electron-phonon and
electron-defect scattering in several theoretical and experimental
studies~\cite{koopmans-jmmm05,koopmans:prl05,faehnle-rapid,cinchetti:prl06,stamm}.
Unlike these papers, we apply, for the first time, an EY mechanism
based exclusively on \emph{electron-electron Coulomb scattering} to
the ultrafast demagnetization in ferromagnetic metals. As a proof of
principle, we demonstrate quantitative agreement for the
demagnetization time and magnetization quenching between time-resolved
magneto-optical Kerr effect measurements on Co and Ni, and the EY
mechanism based on electron-electron scattering.

To resolve the electronic demagnetization dynamics on ultrafast
timescales, we calculate the non-equilibrium, momentum-resolved
multi-band electron dynamics at the level of Boltzmann scattering
integrals, and include the carrier excitation process. We therefore do
not include true electronic correlation effects beyond carrier
scattering nor coherent effects due to the optical excitation
process~\cite{zhang:prl00,Lefkidis-prb2007,Bigot-natphys2009}.

For the description of our general approach, let us assume that the
electronic single-particle energies~$\epsilon^{\mu}_{\vec k}$ and wave
functions $|\mu,\vec{k}\rangle$, where the electronic band index $\mu$
runs over majority- and minority-spin bands, and the vector momentum
is labeled by~$\vec k$, are known from a band structure
calculation. Then the Coulomb and dipole matrix elements can be
calculated and used as an input for dynamical equations for the band
and momentum resolved distribution functions,
$n^{\mu}_{\vec{k}}$. From these, the total magnetization of the system
is obtained by $M=\sum_{\mu=\pm \vec k} s_\mu n^{\mu}_{\vec{k}}$ where
$s_\mu=+1/2$ for majority and $s_\mu=-1/2$ for minority
bands~\footnote{In principle the contributions of the single particle
  states $|\mu,\vec{k}\rangle$ to the total spin is momentum
  dependent~\cite{krauss:prl08}, $s_{\mu}=s_{\mu,\vec{k}}$, and can
be determined from band-structure calculations.}. The equation of
  motion determining the carrier distribution functions has the
  form~\cite{krauss:prl08}
\begin{equation}
\label{ddt}
\frac{\partial n^{\mu}_{\vec k}}{\partial t}  
= \frac{\partial n^{\mu}_{\vec  k}}{\partial t}\Bigr|_{\mathrm{opt}} 
+ \frac{\partial n^{\mu}_{\vec k}}{\partial t}\Bigr|_{\mathrm{e-e}} .
\end{equation}
For the electron-electron Coulomb scattering we use the Boltzmann
equation in the form
\begin{equation}
\label{ddt-n}
\frac{\partial n^{\mu}_{\vec k}}{\partial t}\Bigr|_{\mathrm{e-e}} = \frac{2\pi}{\hbar}
\sum_{\vec l \vec q} 
\sum_{\mu_1 \mu_2 \mu_3}
\bigl |V^{\mu\mu_1}_{\mu_2\mu_3} (\vec k, \vec \ell, \vec q,\omega)\bigr|^2 
 \bigl[ (1-n^{\mu}_{\vec k})n^{\mu_1}_{\vec k + \vec q}(1-n^{\mu_2}_{\vec l + \vec q})n^{\mu_3}_{\vec k} 
-\{(1-n) \leftrightarrow n\} \bigr] 
 \delta(\epsilon^{\mu}_{\vec k} - \epsilon^{\mu_1}_{\vec k +
  \vec q} 
+\epsilon^{\mu_2}_{\vec l + \vec q} -\epsilon^{\mu_3}_{\vec k}).
\end{equation}
where $V$ is the dynamically screened Coulomb potential that depends
on the initial and final states of the two scattering electrons
$|\mu,\vec{k}\rangle \to |\mu_1,\vec{k}+\vec{q}\rangle$ and
$|\mu_2,\vec{l}+\vec{q} \rangle \to |\mu_3,\vec{l}\rangle$, and $\hbar
\omega = \epsilon^{\mu}_{\vec k} - \epsilon^{\mu_1}_{\vec k + \vec
  q}$.  The optical excitation contribution in Eq.~\eqref{ddt} is
calculated by adiabatic elimination of the optical
polarization~\cite{Schaefer-Wegener}
\begin{equation}
\frac{\partial n^{\mu}_{\vec k}}{\partial t}\Bigr|_\mathrm{opt} 
= \sum_{\nu\neq\mu} \bigl|\vec{d}(\vec{k})_{\mu\nu}\cdot \vec{E}\bigr|^2 
(n^{\mu}_{\vec k}-n^{\nu}_{\vec{k}}) 
g(\epsilon^{\nu}_{\vec k}-\epsilon^{\mu}_{\vec k}).
\label{optexc}
\end{equation}
where $\vec{E}$ is the classical electromagnetic field, and the
function $g(\hbar\omega)$ (peaked around the central frequency of the
excitation pulse) models the spread of photon energies which can
induce electronic transitions via the dipole matrix element
$\vec{d}_{\mu\nu}(\vec{k})$ between states $|\nu,k\rangle$ to
$|\mu,k\rangle$. Note that the electron-scattering based version of
the EY mechanism is contained in Eqs.~\eqref{ddt}--\eqref{optexc}, if
the Coulomb matrix elements include the spin-orbit interaction in the
presence of the static lattice, so that scattering transitions change
the average spin of the scattered electrons. The lattice effectively
acts as a sink for the electronic angular momentum, which is ``lost''
from the electronic system by the spin non-conserving scattering
processes described by Eq.~\eqref{ddt-n}. The important Coulomb and
dipole matrix elements can, in principle, be determined from \emph{ab
  initio} treatments~\cite{faehnle-rapid}, and parameter-free results
can be achieved by a dynamical solution of
Eqs.~\eqref{ddt}--\eqref{optexc}. However, due to the numerical
complexity of the $k$-resolved Boltzmann scattering
integral~\eqref{ddt-n}, we use a simplified model that contains
parameters. We stress that the parameters and simplifications
introduced below can be eliminated by using input from \emph{ab
  initio} methods. We approximate the energy bands as spherically
symmetric, $\epsilon^{\mu}_{\vec k} = \epsilon^{\mu}_{|k|}$, and the
screened Coulomb interaction as
\begin{equation}
\label{Coulomb-approx}
V^{\mu\mu_1}_{\mu_2\mu_3} (\vec k, \vec \ell, \vec q,\omega)
= f_{\mu}^{\mu_1} f_{\mu_2}^{\mu_3} \frac{v(q)}{\varepsilon(q,\omega)}
\end{equation}
where 
\begin{equation}
f_{\mu}^{\mu'} =
\begin{cases} 
1 \text{ if } \mu = \mu' \\
\alpha \text{ if } \mu\neq\mu' 
\end{cases}
\end{equation}
The parameter $\alpha$ is roughly comparable
to the $\alpha$ parameter introduced by Yafet~\cite{yafet-ssp63} and
calculated recently for metallic ferromagnets~\cite{faehnle-rapid}. In
Eq.~\eqref{Coulomb-approx}, $v(q)$ is the bare Coulomb potential, and
$\varepsilon^{-1}(q,\omega)$ the dynamical inverse dielectric function. For
$\vec{k}$, $\vec{l}$, and $\vec{q}$ dependent factors $f$, this form
can be shown to be valid if there are no short-range contributions to
the Coulomb interaction~\cite{pikus-bir:71}, and we use this as an
approximate explicit expression for the Coulomb matrix element of
metals. Important dynamical screening effects are included via the
Lindhard dielectric function $\varepsilon(\vec q,\omega)$. In
semiconductors, it has recently been demonstrated that an approach
closely related~\cite{krauss:prl08} to the one presented here leads to
a parameter free agreement for the spin dynamics in theory and
experiment~\cite{hilton-tang} because quite accurate wavefunctions can
be obtained using $\vec k \cdot \vec p$ theory. Finally, we assume
that optical excitation connects only majority bands and minority
bands with each other, respectively, and we approximate the strength
of the optical dipole matrix elements by a momentum and
band-independent constant~$\vec{d}$. Although this is a drastic
oversimplification, especially in view of the hybridization between
$s$ and $p$ bands, it is in the same spirit as the approximations
introduced for the band structure and the Coulomb interaction: The
dependence on the electron vector momentum should either be included
in all these quantities, or modeled in a way that introduces the least
amount of parameters in the model. In this paper, the electronic
excitation after the optical pulse is therefore determined by the band
structure, the central photon energy and width of pump pulse, as
well as the fluence. These quantities are used as input for the
numerical calculations.

Finally, we include the equilibration of the electronic system with
the lattice via a relaxation time approximation 
\begin{equation}
\frac{\partial n^{\mu}_{\vec k}}{\partial t}\Bigr|_\mathrm{therm} 
= \frac{n^{\mu}_{\vec k}- f^{\mu}_{\vec k}}{\tau_{\mathrm{phon}}}
\end{equation}
 in Eq.~\eqref{ddt}. Here,
$f^{\mu}_{\vec k}$ denotes the Fermi-Dirac distribution of electrons
in band~$\mu$ at lattice temperature. Note that in our model the
demagnetization solely occurs due to Coulomb scattering, and that the
electron-phonon interaction only leads to thermal equilibration, which
actually restores the ground-state magnetization. The equilibration
times~$\tau_\mathrm{phon,Ni} = 25$\,ps and $\tau_\mathrm{phon,Co} =
5$\,ps are extracted from experiment by fitting the remagnetization
dynamics. This assumption is based on the experimental observation
that the timescale for energy equilibration due to electron-phonon
interaction, which includes the effect of heat diffusion in apragmatic
way, is typically longer than the demagnetization time.

On the experimental side, a variety of techniques are available to
excite and detect electron-spin dynamics~\cite{ma:prl97}. Here, we
apply an all-optical strategy to trace the spin dynamics on
femtosecond timescales. By means of the time-resolved magneto-optical
Kerr effect (TR-MOKE) in the longitudinal configuration we excite the
ferromagnet by an ultrafast optical pump pulse, and monitor the
material response by a delayed and modified optical replica (probe
pulse). The femtosecond pulses are generated by a Ti:Sapphire
multipass amplifier with 1\,kHz repetition rate. We use s-polarized
50\,fs, 800\,nm pump pulses at normal incidence, and s-polarized
50\,fs, 400\,nm probe pulses under 45$^{\circ}$. The samples are thin
polycrystalline ferromagnetic layers: a 15\,nm cobalt film deposited
on MgO by dc-sputtering, and a 15\,nm Ni film deposited on Si by
electron-beam evaporation. The Ni film is capped by a 3\,nm Ti layer;
another 3\,nm Ti layer acts as an adhesion promoter between the Ni
film and the substrate.

\begin{figure}[t]
\includegraphics[width=0.45\textwidth]{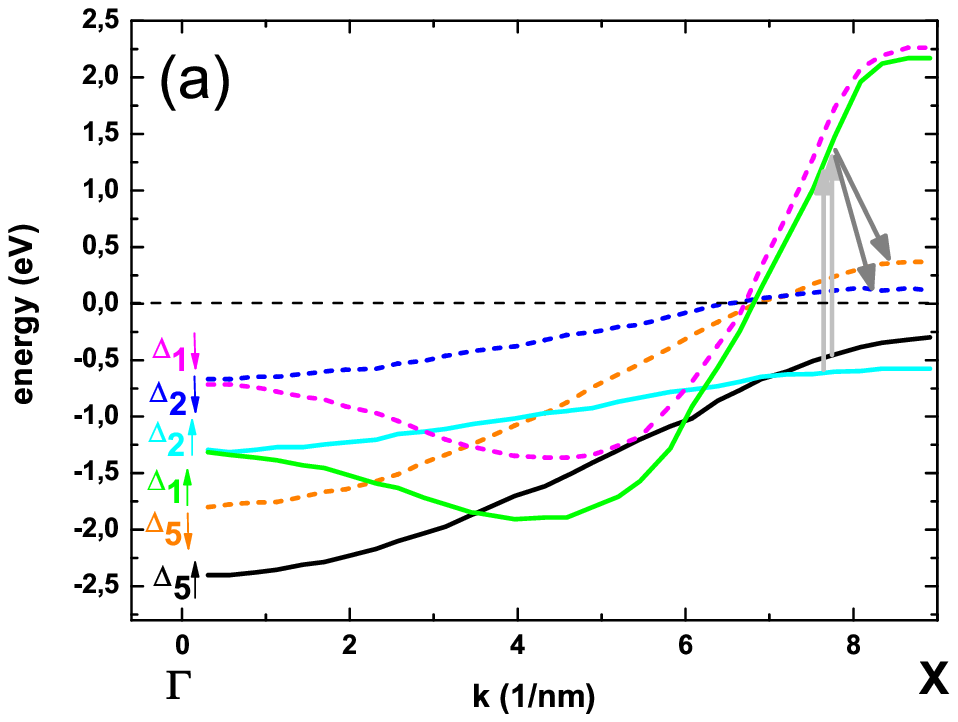}
\includegraphics[width=0.45\textwidth]{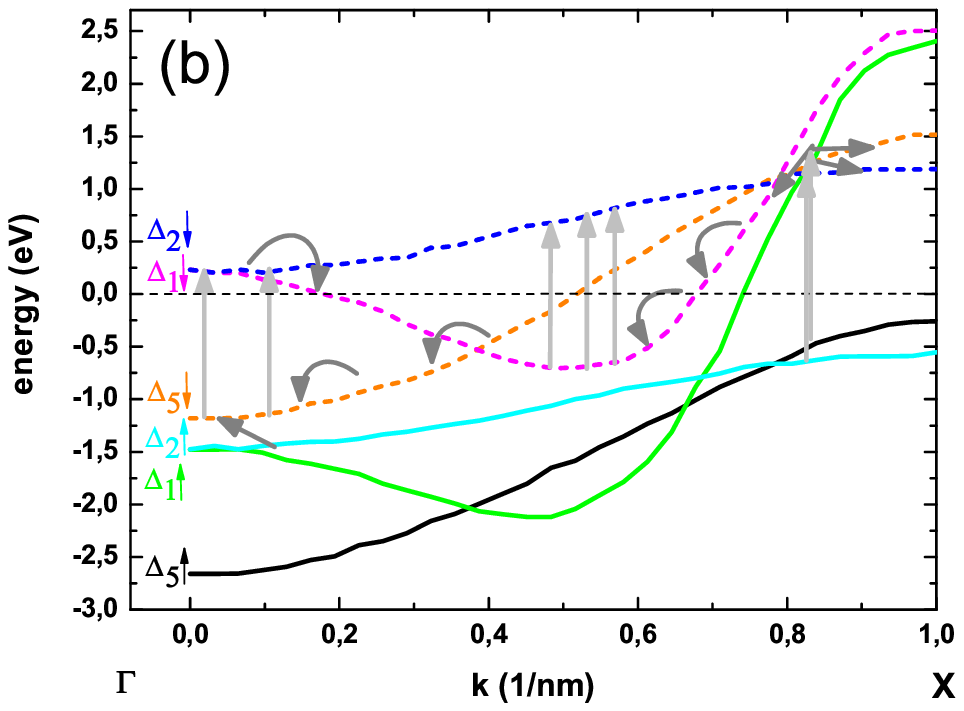}
\caption{Band structure $\epsilon^{\mu}_{k}$ of Ni (a) and Co (b) for
  majority (solid lines) and minority (dashed lines) electrons in
  $\Gamma$-$X$ direction~\cite{landolt-boernstein}.  The Fermi energy
  $E_{\mathrm{F}}$ is set to zero. The arrows indicate a typical
  ultrafast demagnetization scenario: Electrons are excited by an
  ultrashort laser pulse (vertical light grey arrows), which does not
  change the total magnetization. They relax via intra- and inter-band
  scattering (grey arrows). The latter scattering process leads to
  depolarization of the electrons. }
\label{figure1}
\end{figure}

Qualitatively, the ultrafast demagnetization occurs in our model in
the following way.  The electronic distributions in the unexcited
ferromagnet are assumed to be Fermi-Dirac distributions determined by
the lattice temperature and the band structure. The majority and
minority energy dispersions are spin split so that a non-zero
magnetization exists in equilibrium. The ultrafast optical excitation
process creates nonequilibrium electrons in bands accessible by the
pump photon energy, and the electrons undergo intra\-band and
interband Coulomb scattering processes. The driving force for the
demagnetization are interband scattering processes between the
optically excited electrons, which lead to the redistribution of
electrons from majority to minority bands as long as the optically
excited electrons are spin polarized. Remagnetization occurs due to
equilibration at lattice temperature, because the groundstate
magnetization is restored when the electrons settle down in the band
minima.

%
%

\begin{figure}[tb]
\rotatebox{270}{\includegraphics[height=0.9\textwidth]{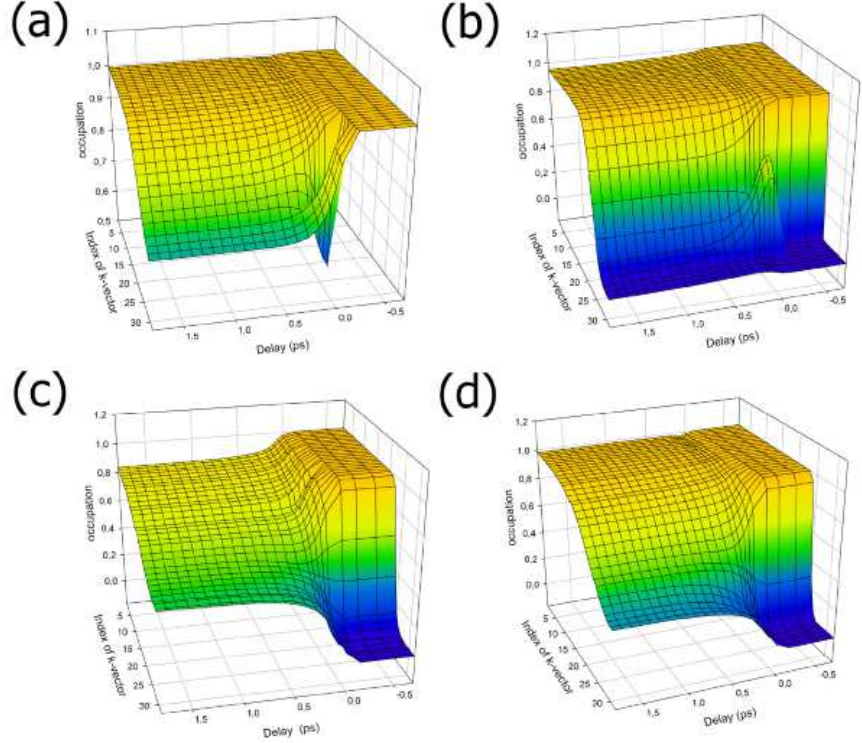}}
\caption{Dynamical distribution functions $n^{\mu}_{\vec{k}}$ for four
  different bands. Electrons are excited from bands
  $\Delta_2^\uparrow$ and $\Delta_5^\uparrow$ (a) into band
  $\Delta_1^\uparrow$ (b). Ultrafast demagnetization occurs by
  scattering into bands $\Delta_2^\downarrow$ (c) and
  $\Delta_5^\downarrow$ (d). ($k= 8.9$\,nm$^{-1}\times$ the index in
  the figures)}
\label{figure2}
\end{figure}

For the numerical calculations we use as input for
$\epsilon^{\mu}_{|k|}$ a KKR-DFT-result~\cite{landolt-boernstein} for
the $\Gamma$-$X$ direction, which is then used for the whole Brillouin
zone as if the band structure were spherically symmetric. These
dispersions are plotted in Fig~\ref{figure1}. For an experimental and
theoretical study of Co and Ni band structures that also discusses the
nomenclature of the bands, see Ref.~\onlinecite{Tobin-prb99}. The
exciting laser pulse has a typical full-width at half-maximum of
50\,fs, photon energy of 1.55\,eV and the fluence is numerically
adjusted to be in qualitative agreement with an estimate of the
absorption and the observed magnetization quenching.  The pump pulse
excites electrons into initially empty states above the Fermi
energy~$E_{\mathrm{F}}$, as modeled by Eq.~\eqref{optexc}. Some
numerical results obtained from Eq.~\eqref{ddt} for the time- and
momentum-resolved electron occupation for Ni are shown in
Fig.~\ref{figure2}. Since the distribution functions contain all the
information on the dynamics on the single-particle level, we use them,
together with the band structure shown in Fig.~\ref{figure1}, to
discuss the demagnetization scenario for Ni. Optical excitation by the
ultrashort 1.55\,eV, excitation pulse is only possible for transitions
from the $\Delta_2^\uparrow$ and $\Delta_5^\uparrow$ bands to the
$\Delta_1^\uparrow$ (light grey arrows in
Fig.~\ref{figure1}). Fig.~\ref{figure2}(a) shows the nonequilibrium
distributions created by the pump pulse. During and after the optical
excitation of electrons in $\Delta_1^{\uparrow}$, electron-electron
scattering processes redistribute the carriers in and between the
bands. In Figs.~\ref{figure2} (c) and (d), the increasing number of
electrons in $\Delta_2^\downarrow$ and $\Delta_5^\downarrow$ for
positive time delays above $k=7$\,nm$^{-1}$ illustrates the dominant
scattering pathways. Note that the scattering of electrons from
majority to minority bands reduces the overall magnetization, as the
electronic contribution to the expectation value of the spin is
altered. The processes responsible for the ultrafast loss of magnetic
order in Ni start more than 1\,eV above the Fermi energy
$E_{\mathrm{F}}$) and mainly take place near the
$X$-point. Remarkably, the demagnetization is almost completely
dominated by the two transitions mentioned above, with the
nonequilibrium scattering dynamics taking place over more than
100\,fs, cf.~Figs. \ref{figure2} (c) and (d). Electrons that are not
scattered out of band $\Delta_1^\uparrow$ at high energies above the
Fermi level accumulate in states close to the Fermi-level because
energy- and momentum conservation requirements make out-scattering
processes inefficient. For completeness, we mention that band
$\Delta_1^\downarrow$ does not play an important role in the
demagnetization dynamics of Ni.  An analysis of the electronic
occupation in the different bands for Co along the same lines leads to
the scenario as depicted in Fig.~\ref{figure1}(b).

\begin{figure}[t]
\includegraphics[width=0.6\textwidth]{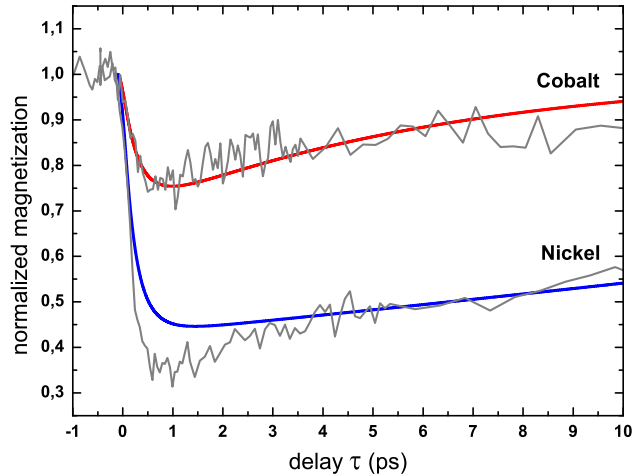}
\caption{Normalized TR-MOKE rotation (grey line) and calculated
  magnetization dynamics for Ni (blue line) and Co (red line).
  Assuming the same laser fluence, the choice of parameters
  $\alpha_\mathrm{Co}=0.15$ and $\alpha_\mathrm{Ni}=0.3$ yields good
  agreement between theory and experiment. The demagnetization times
  are $\tau_{\mathrm{Co}}= 215$\,fs and $\tau_{\mathrm{Ni}}=
  200$\,fs. In the calculation, the stronger quenching of the
  magnetization for Ni is due to band structure effects and not very
  sensitive to the numerical value of $\alpha$.}
\label{figure3}
\end{figure}

In Figure~\ref{figure3}, we plot both the signal obtained from TR-MOKE
measurements and the calculated signal for Ni and Co. The
magnetization quenching for Ni is significantly stronger than for
Co. In our model this is explained by the band structure in
combination with the optical excitation process, which yields a more
efficient carrier excitation with 800\,nm photons: in Ni mainly
majority electrons are optically excited (see Fig.~\ref{figure1}) so
that all interband scattering processes above $E_{\mathrm{F}}$ lead to
demagnetization. To obtain quantitative agreement between theory and
experiment, we assume the same laser fluence for both materials, and
use the Elliott-Yafet factor $\alpha$, introduced
in~Eq.~\eqref{Coulomb-approx}, as a single fit parameter. We obtain
$\alpha_{\mathrm{Co}}=0.15$ and $\alpha_{\mathrm{Ni}}=0.3$ together
with the demagnetization times of $\tau_{\mathrm{Co}}= 215$\,fs and
$\tau_{\mathrm{Ni}}= 200$\,fs. The general trend
$\alpha_\mathrm{Co}<\alpha_\mathrm{Ni}$ and the order of magnitude
compare well with recent \emph{ab initio} results for the $\alpha$
parameter~\cite{faehnle-rapid}. The results in
Ref.~\onlinecite{faehnle-rapid} provide only a qualitative check for
our fit parameters, because the \emph{ab initio} results depend on the
band-structure region, over which the wave-function coefficients are
averaged. Last, but not least, we stress that the band structure
properties influence the microscopic dynamics sufficiently strongly to
make it impossible to fit the Co measurements in Fig.~\ref{figure3}
using the Ni band-structure and, vice versa. 


In conclusion, we applied, an Elliott-Yafet mechanism based on
electron-electron Coulomb scattering for the explanation of the
optically induced ultrafast demagnetization in the ferromagnetic
transition metals Ni and Co. The electronic demagnetization in our
model occurs through the Coulomb interaction, which is spin-diagonal
for free electrons, in the presence of the spin-orbit
interaction. Modeling the optical excitation process and the
scattering dynamics by Boltzmann scattering integrals for the
momentum-dependent dynamical distributions functions in the various
bands, we described a model that can make parameter-free predictions
from input based \emph{ab initio} band structure results. In this
paper, we made simplifying assumptions for the Coulomb and dipole
matrix elements, which led to the introduction of two parameters, the
EY-parameter $\alpha$ and the fluence, which are fixed by comparison
with experiment. Even though the origin of ferromagnetism is an
interaction effect of many electrons, the good agreement between
theory and experiment presents evidence that the ultrafast
demagnetization dynamics in ferromagnets can be understood in a
single-particle picture including electron-electron scattering and the
spin-orbit interaction. Unlike other EY based models, we do not take
into account the electron-phonon interaction for the demagnetization
dynamics. Our goal was to present a proof-of-principle that ultrafast
demagnetization can occur due to an EY mechanism based on
electron-electron scattering. The question whether the
electron-electron or electron-phonon interaction predominates should
be investigated further. Theoretically this can be achieved using the
tools developed in this paper with the help of \emph{ab initio} band
structure calculations. Experimental strategies to separate the two
contributions should be developed concurrently.

\begin{acknowledgments}
We acknowledge support from the Graduiertenkolleg~792 ``Nonlinear
Optics and Ultrafast Processes'' and SPP~1133, and from the NIC
J\"{u}lich through a CPU time grant. We thank Christoph D\"oring for
sample preparation, and M.~F\"ahnle (MPI Stuttgart) for helpful
discussions.
\end{acknowledgments}

\bibliographystyle{apsrev}
\bibliography{metals}

\end{document}